\documentclass{article}
\usepackage{macros/spconf,amsmath,graphicx,enumitem}
\usepackage{xcolor}
\usepackage{microtype}
\usepackage{subfigure}
\usepackage{hyperref}
\usepackage{cite}

\usepackage{amsthm}

\usepackage{amssymb}
\usepackage{amsfonts}
\usepackage{mathrsfs}
\usepackage{xspace}
\usepackage{bm}
\usepackage{upgreek}

\newcommand{\safemath}[2]{\newcommand{#1}{\ensuremath{#2}\xspace}}

\safemath{\bma}{\mathbf{a}}
\safemath{\bmb}{\mathbf{b}}
\safemath{\bmc}{\mathbf{c}}
\safemath{\bmd}{\mathbf{d}}
\safemath{\bme}{\mathbf{e}}
\safemath{\bmf}{\mathbf{f}}
\safemath{\bmg}{\mathbf{g}}
\safemath{\bmh}{\mathbf{h}}
\safemath{\bmi}{\mathbf{i}}
\safemath{\bmj}{\mathbf{j}}
\safemath{\bmk}{\mathbf{k}}
\safemath{\bml}{\mathbf{l}}
\safemath{\bmm}{\mathbf{m}}
\safemath{\bmn}{\mathbf{n}}
\safemath{\bmo}{\mathbf{o}}
\safemath{\bmp}{\mathbf{p}}
\safemath{\bmq}{\mathbf{q}}
\safemath{\bmr}{\mathbf{r}}
\safemath{\bms}{\mathbf{s}}
\safemath{\bmt}{\mathbf{t}}
\safemath{\bmu}{\mathbf{u}}
\safemath{\bmv}{\mathbf{v}}
\safemath{\bmw}{\mathbf{w}}
\safemath{\bmx}{\mathbf{x}}
\safemath{\bmy}{\mathbf{y}}
\safemath{\bmz}{\mathbf{z}}
\safemath{\bmzero}{\mathbf{0}}
\safemath{\bmone}{\mathbf{1}}

\bmdefine{\biad}{a}
\bmdefine{\bibd}{b}
\bmdefine{\bicd}{c}
\bmdefine{\bidd}{d}
\bmdefine{\bied}{e}
\bmdefine{\bifd}{f}
\bmdefine{\bigd}{g}
\bmdefine{\bihd}{h}
\bmdefine{\biid}{i}
\bmdefine{\bijd}{j}
\bmdefine{\bikd}{k}
\bmdefine{\bild}{l}
\bmdefine{\bimd}{m}
\bmdefine{\bind}{n}
\bmdefine{\biod}{o}
\bmdefine{\bipd}{p}
\bmdefine{\biqd}{q}
\bmdefine{\bird}{r}
\bmdefine{\bisd}{s}
\bmdefine{\bitd}{t}
\bmdefine{\biud}{u}
\bmdefine{\bivd}{v}
\bmdefine{\biwd}{w}
\bmdefine{\bixd}{x}
\bmdefine{\biyd}{y}
\bmdefine{\bizd}{z}

\bmdefine{\bixid}{\xi}
\bmdefine{\bilambdad}{\lambda}
\bmdefine{\bimud}{\mu}
\bmdefine{\bithetad}{\theta}
\bmdefine{\biphid}{\phi}
\bmdefine{\bideltad}{\delta}

\safemath{\bmia}{\biad}
\safemath{\bmib}{\bibd}
\safemath{\bmic}{\bicd}
\safemath{\bmid}{\bidd}
\safemath{\bmie}{\bied}
\safemath{\bmif}{\bifd}
\safemath{\bmig}{\bigd}
\safemath{\bmih}{\bihd}
\safemath{\bmii}{\biid}
\safemath{\bmij}{\bijd}
\safemath{\bmik}{\bikd}
\safemath{\bmil}{\bild}
\safemath{\bmim}{\bimd}
\safemath{\bmin}{\bind}
\safemath{\bmio}{\biod}
\safemath{\bmip}{\bipd}
\safemath{\bmiq}{\biqd}
\safemath{\bmir}{\bird}
\safemath{\bmis}{\bisd}
\safemath{\bmit}{\bitd}
\safemath{\bmiu}{\biud}
\safemath{\bmiv}{\bivd}
\safemath{\bmiw}{\biwd}
\safemath{\bmix}{\bixd}
\safemath{\bmiy}{\biyd}
\safemath{\bmiz}{\bizd}

\safemath{\bmxi}{\bixid}
\safemath{\bmlambda}{\bilambdad}
\safemath{\bmmu}{\bimud}
\safemath{\bmtheta}{\bithetad}
\safemath{\bmphi}{\biphid}
\safemath{\bmdelta}{\bideltad}

\safemath{\bA}{\mathbf{A}}
\safemath{\bB}{\mathbf{B}}
\safemath{\bC}{\mathbf{C}}
\safemath{\bD}{\mathbf{D}}
\safemath{\bE}{\mathbf{E}}
\safemath{\bF}{\mathbf{F}}
\safemath{\bG}{\mathbf{G}}
\safemath{\bH}{\mathbf{H}}
\safemath{\bI}{\mathbf{I}}
\safemath{\bJ}{\mathbf{J}}
\safemath{\bK}{\mathbf{K}}
\safemath{\bL}{\mathbf{L}}
\safemath{\bM}{\mathbf{M}}
\safemath{\bN}{\mathbf{N}}
\safemath{\bO}{\mathbf{O}}
\safemath{\bP}{\mathbf{P}}
\safemath{\bQ}{\mathbf{Q}}
\safemath{\bR}{\mathbf{R}}
\safemath{\bS}{\mathbf{S}}
\safemath{\bT}{\mathbf{T}}
\safemath{\bU}{\mathbf{U}}
\safemath{\bV}{\mathbf{V}}
\safemath{\bW}{\mathbf{W}}
\safemath{\bX}{\mathbf{X}}
\safemath{\bY}{\mathbf{Y}}
\safemath{\bZ}{\mathbf{Z}}

\safemath{\bZero}{\mathbf{0}}
\safemath{\bOne}{\mathbf{1}}
\safemath{\bDelta}{\mathbf{\Delta}}
\safemath{\bLambda}{\mathbf{\UpLambda}}
\safemath{\bPhi}{\mathbf{\Upphi}}
\safemath{\bSigma}{\mathbf{\Upsigma}}
\safemath{\bOmega}{\mathbf{\Upomega}}
\safemath{\bTheta}{\mathbf{\Uptheta}}

\bmdefine{\biAd}{A}
\bmdefine{\biBd}{B}
\bmdefine{\biCd}{C}
\bmdefine{\biDd}{D}
\bmdefine{\biEd}{E}
\bmdefine{\biFd}{F}
\bmdefine{\biGd}{G}
\bmdefine{\biHd}{H}
\bmdefine{\biId}{I}
\bmdefine{\biJd}{J}
\bmdefine{\biKd}{K}
\bmdefine{\biLd}{L}
\bmdefine{\biMd}{M}
\bmdefine{\biOd}{N}
\bmdefine{\biPd}{O}
\bmdefine{\biQd}{P}
\bmdefine{\biRd}{R}
\bmdefine{\biSd}{S}
\bmdefine{\biTd}{T}
\bmdefine{\biUd}{U}
\bmdefine{\biVd}{V}
\bmdefine{\biWd}{W}
\bmdefine{\biXd}{X}
\bmdefine{\biYd}{Y}
\bmdefine{\biZd}{Z}

\bmdefine{\biDelta}{\Delta}
\bmdefine{\biLambda}{\Lambda}
\bmdefine{\biPhi}{\Phi}
\bmdefine{\biSigma}{\Sigma}
\bmdefine{\biOmega}{\Omega}
\bmdefine{\biTheta}{\Theta}

\safemath{\bimA}{\biAd}
\safemath{\bimB}{\biBd}
\safemath{\bimC}{\biCd}
\safemath{\bimD}{\biDd}
\safemath{\bimE}{\biEd}
\safemath{\bimF}{\biFd}
\safemath{\bimG}{\biGd}
\safemath{\bimH}{\biHd}
\safemath{\bimI}{\biId}
\safemath{\bimJ}{\biJd}
\safemath{\bimK}{\biKd}
\safemath{\bimL}{\biLd}
\safemath{\bimM}{\biMd}
\safemath{\bimN}{\biNd}
\safemath{\bimO}{\biOd}
\safemath{\bimP}{\biPd}
\safemath{\bimQ}{\biQd}
\safemath{\bimR}{\biRd}
\safemath{\bimS}{\biSd}
\safemath{\bimT}{\biTd}
\safemath{\bimU}{\biUd}
\safemath{\bimV}{\biVd}
\safemath{\bimW}{\biWd}
\safemath{\bimX}{\biXd}
\safemath{\bimY}{\biYd}
\safemath{\bimZ}{\biZd}

\safemath{\bimDelta}{\biDelta}
\safemath{\bimLambda}{\biLambda}
\safemath{\bimPhi}{\biPhi}
\safemath{\bimSigma}{\biSigma}
\safemath{\bimOmega}{\biOmega}
\safemath{\bimTheta}{\biTheta}

\safemath{\setA}{\mathcal{A}}
\safemath{\setB}{\mathcal{B}}
\safemath{\setC}{\mathcal{C}}
\safemath{\setD}{\mathcal{D}}
\safemath{\setE}{\mathcal{E}}
\safemath{\setF}{\mathcal{F}}
\safemath{\setG}{\mathcal{G}}
\safemath{\setH}{\mathcal{H}}
\safemath{\setI}{\mathcal{I}}
\safemath{\setJ}{\mathcal{J}}
\safemath{\setK}{\mathcal{K}}
\safemath{\setL}{\mathcal{L}}
\safemath{\setM}{\mathcal{M}}
\safemath{\setN}{\mathcal{N}}
\safemath{\setO}{\mathcal{O}}
\safemath{\setP}{\mathcal{P}}
\safemath{\setQ}{\mathcal{Q}}
\safemath{\setR}{\mathcal{R}}
\safemath{\setS}{\mathcal{S}}
\safemath{\setT}{\mathcal{T}}
\safemath{\setU}{\mathcal{U}}
\safemath{\setV}{\mathcal{V}}
\safemath{\setW}{\mathcal{W}}
\safemath{\setX}{\mathcal{X}}
\safemath{\setY}{\mathcal{Y}}
\safemath{\setZ}{\mathcal{Z}}
\safemath{\emptySet}{\varnothing}

\safemath{\colA}{\mathscr{A}}
\safemath{\colB}{\mathscr{B}}
\safemath{\colC}{\mathscr{C}}
\safemath{\colD}{\mathscr{D}}
\safemath{\colE}{\mathscr{E}}
\safemath{\colF}{\mathscr{F}}
\safemath{\colG}{\mathscr{G}}
\safemath{\colH}{\mathscr{H}}
\safemath{\colI}{\mathscr{I}}
\safemath{\colJ}{\mathscr{J}}
\safemath{\colK}{\mathscr{K}}
\safemath{\colL}{\mathscr{L}}
\safemath{\colM}{\mathscr{M}}
\safemath{\colN}{\mathscr{N}}
\safemath{\colO}{\mathscr{O}}
\safemath{\colP}{\mathscr{P}}
\safemath{\colQ}{\mathscr{Q}}
\safemath{\colR}{\mathscr{R}}
\safemath{\colS}{\mathscr{S}}
\safemath{\colT}{\mathscr{T}}
\safemath{\colU}{\mathscr{U}}
\safemath{\colV}{\mathscr{V}}
\safemath{\colW}{\mathscr{W}}
\safemath{\colX}{\mathscr{X}}
\safemath{\colY}{\mathscr{Y}}
\safemath{\colZ}{\mathscr{Z}}

\safemath{\opA}{\mathbb{A}}
\safemath{\opB}{\mathbb{B}}
\safemath{\opC}{\mathbb{C}}
\safemath{\opD}{\mathbb{D}}
\safemath{\opE}{\mathbb{E}}
\safemath{\opF}{\mathbb{F}}
\safemath{\opG}{\mathbb{G}}
\safemath{\opH}{\mathbb{H}}
\safemath{\opI}{\mathbb{I}}
\safemath{\opJ}{\mathbb{J}}
\safemath{\opK}{\mathbb{K}}
\safemath{\opL}{\mathbb{L}}
\safemath{\opM}{\mathbb{M}}
\safemath{\opN}{\mathbb{N}}
\safemath{\opO}{\mathbb{O}}
\safemath{\opP}{\mathbb{P}}
\safemath{\opQ}{\mathbb{Q}}
\safemath{\opR}{\mathbb{R}}
\safemath{\opS}{\mathbb{S}}
\safemath{\opT}{\mathbb{T}}
\safemath{\opU}{\mathbb{U}}
\safemath{\opV}{\mathbb{V}}
\safemath{\opW}{\mathbb{W}}
\safemath{\opX}{\mathbb{X}}
\safemath{\opY}{\mathbb{Y}}
\safemath{\opZ}{\mathbb{Z}}
\safemath{\opZero}{\mathbb{O}}
\safemath{\identityop}{\opI}

\safemath{\veca}{\bma}
\safemath{\vecb}{\bmb}
\safemath{\vecc}{\bmc}
\safemath{\vecd}{\bmd}
\safemath{\vece}{\bme}
\safemath{\vecf}{\bmf}
\safemath{\vecg}{\bmg}
\safemath{\vech}{\bmh}
\safemath{\veci}{\bmi}
\safemath{\vecj}{\bmj}
\safemath{\veck}{\bmk}
\safemath{\vecl}{\bml}
\safemath{\vecm}{\bmm}
\safemath{\vecn}{\bmn}
\safemath{\veco}{\bmo}
\safemath{\vecp}{\bmp}
\safemath{\vecq}{\bmq}
\safemath{\vecr}{\bmr}
\safemath{\vecs}{\bms}
\safemath{\vect}{\bmt}
\safemath{\vecu}{\bmu}
\safemath{\vecv}{\bmv}
\safemath{\vecw}{\bmw}
\safemath{\vecx}{\bmx}
\safemath{\vecy}{\bmy}
\safemath{\vecz}{\bmz}

\safemath{\veczero}{\bmzero}
\safemath{\vecone}{\bmone}
\safemath{\vecxi}{\bmxi}
\safemath{\veclambda}{\bmlambda}
\safemath{\vecmu}{\bmmu}
\safemath{\vectheta}{\bmtheta}
\safemath{\vecphi}{\bmphi}
\safemath{\vecdelta}{\bmdelta}

\safemath{\matA}{\bA}
\safemath{\matB}{\bB}
\safemath{\matC}{\bC}
\safemath{\matD}{\bD}
\safemath{\matE}{\bE}
\safemath{\matF}{\bF}
\safemath{\matG}{\bG}
\safemath{\matH}{\bH}
\safemath{\matI}{\bI}
\safemath{\matJ}{\bJ}
\safemath{\matK}{\bK}
\safemath{\matL}{\bL}
\safemath{\matM}{\bM}
\safemath{\matN}{\bN}
\safemath{\matO}{\bO}
\safemath{\matP}{\bP}
\safemath{\matQ}{\bQ}
\safemath{\matR}{\bR}
\safemath{\matS}{\bS}
\safemath{\matT}{\bT}
\safemath{\matU}{\bU}
\safemath{\matV}{\bV}
\safemath{\matW}{\bW}
\safemath{\matX}{\bX}
\safemath{\matY}{\bY}
\safemath{\matZ}{\bZ}
\safemath{\matzero}{\bmzero}

\safemath{\matDelta}{\bDelta}
\safemath{\matLambda}{\bLambda}
\safemath{\matPhi}{\bPhi}
\safemath{\matSigma}{\bSigma}
\safemath{\matOmega}{\bOmega}
\safemath{\matTheta}{\bTheta}

\safemath{\matidentity}{\matI}
\safemath{\matone}{\matO}

\safemath{\rnda}{A}
\safemath{\rndb}{B}
\safemath{\rndc}{C}
\safemath{\rndd}{D}
\safemath{\rnde}{E}
\safemath{\rndf}{F}
\safemath{\rndg}{G}
\safemath{\rndh}{H}
\safemath{\rndi}{I}
\safemath{\rndj}{J}
\safemath{\rndk}{K}
\safemath{\rndl}{L}
\safemath{\rndm}{M}
\safemath{\rndn}{N}
\safemath{\rndo}{O}
\safemath{\rndp}{P}
\safemath{\rndq}{Q}
\safemath{\rndr}{R}
\safemath{\rnds}{S}
\safemath{\rndt}{T}
\safemath{\rndu}{U}
\safemath{\rndv}{V}
\safemath{\rndw}{W}
\safemath{\rndx}{X}
\safemath{\rndy}{Y}
\safemath{\rndz}{Z}

\safemath{\rveca}{\bimA}
\safemath{\rvecb}{\bimB}
\safemath{\rvecc}{\bimC}
\safemath{\rvecd}{\bimD}
\safemath{\rvece}{\bimE}
\safemath{\rvecf}{\bimF}
\safemath{\rvecg}{\bimG}
\safemath{\rvech}{\bimH}
\safemath{\rveci}{\bimI}
\safemath{\rvecj}{\bimJ}
\safemath{\rveck}{\bimK}
\safemath{\rvecl}{\bimL}
\safemath{\rvecm}{\bimM}
\safemath{\rvecn}{\bimN}
\safemath{\rveco}{\bomO}
\safemath{\rvecp}{\bimP}
\safemath{\rvecq}{\bimQ}
\safemath{\rvecr}{\bimR}
\safemath{\rvecs}{\bimS}
\safemath{\rvect}{\bimT}
\safemath{\rvecu}{\bimU}
\safemath{\rvecv}{\bimV}
\safemath{\rvecw}{\bimW}
\safemath{\rvecx}{\bimX}
\safemath{\rvecy}{\bimY}
\safemath{\rvecz}{\bimZ}

\safemath{\rvecxi}{\bmxi}
\safemath{\rveclambda}{\bmlambda}
\safemath{\rvecmu}{\bmmu}
\safemath{\rvectheta}{\bmtheta}
\safemath{\rvecphi}{\bmphi}

\safemath{\rmatA}{\bimA}
\safemath{\rmatB}{\bimB}
\safemath{\rmatC}{\bimC}
\safemath{\rmatD}{\bimD}
\safemath{\rmatE}{\bimE}
\safemath{\rmatF}{\bimF}
\safemath{\rmatG}{\bimG}
\safemath{\rmatH}{\bimH}
\safemath{\rmatI}{\bimI}
\safemath{\rmatJ}{\bimJ}
\safemath{\rmatK}{\bimK}
\safemath{\rmatL}{\bimL}
\safemath{\rmatM}{\bimM}
\safemath{\rmatN}{\bimN}
\safemath{\rmatO}{\bimO}
\safemath{\rmatP}{\bimP}
\safemath{\rmatQ}{\bimQ}
\safemath{\rmatR}{\bimR}
\safemath{\rmatS}{\bimS}
\safemath{\rmatT}{\bimT}
\safemath{\rmatU}{\bimU}
\safemath{\rmatV}{\bimV}
\safemath{\rmatW}{\bimW}
\safemath{\rmatX}{\bimX}
\safemath{\rmatY}{\bimY}
\safemath{\rmatZ}{\bimZ}

\safemath{\rmatDelta}{\bimDelta}
\safemath{\rmatLambda}{\bimLambda}
\safemath{\rmatPhi}{\bimPhi}
\safemath{\rmatSigma}{\bimSigma}
\safemath{\rmatOmega}{\bimOmega}
\safemath{\rmatTheta}{\bimTheta}

\usepackage{amssymb}
\usepackage{amsfonts}
\usepackage{mathrsfs}
\usepackage{xspace}
\usepackage{bm}
\usepackage{fancyref}
\usepackage{textcomp}

\usepackage{multirow}
\usepackage{stmaryrd}

\newenvironment{textbmatrix}{	\setlength{\arraycolsep}{2.5pt}%
								\big[\begin{matrix}}{\end{matrix}\big]%
								\raisebox{0.08ex}{\vphantom{M}}}

\def\be{\begin{equation}}
\def\ee{\end{equation}}
\def\een{\nonumber \end{equation}}
\def\mat{\begin{bmatrix}}
\def\emat{\end{bmatrix}}
\def\btm{\begin{textbmatrix}}
\def\etm{\end{textbmatrix}}

\def\ba#1\ea{\begin{align}#1\end{align}}
\def\bas#1\eas{\begin{align*}#1\end{align*}}
\def\bs#1\es{\begin{split}#1\end{split}} 
\def\bg#1\eg{\begin{gather}#1\end{gather}}
\def\bml#1\eml{\begin{multline}#1\end{multline}}
\def\bi#1\ei{\begin{itemize}#1\end{itemize}}

\newcommand{\lefto}{\mathopen{}\left}

\DeclareMathOperator*{\argmin}{arg\;min}		
\DeclareMathOperator{\Exop}{\opE}			

\newcommand{\Ex}[2]{\ensuremath{\Exop_{#1}\lefto[#2\right]}} 	

\newcommand{\vecnorm}[1]{\lefto\lVert#1\right\rVert}		

\safemath{\dirac}{\delta}					
\safemath{\krond}{\dirac}					

\safemath{\upto}{\uparrow}
\safemath{\downto}{\downarrow}
\safemath{\iu}{j}							
\safemath{\ev}{\lambda}						
\safemath{\hilseqspace}{l^{2}}				
\newcommand{\banachfunspace}[1]{\setL^{#1}}	
\safemath{\hilfunspace}{\banachfunspace{2}}	

\safemath{\SNR}{\textsf{SNR}} 				
\safemath{\PAR}{\textsf{PAR}} 				
\safemath{\No}{N_0}							
\safemath{\Es}{E_s}							
\safemath{\Eb}{E_b}							
\safemath{\EbNo}{\frac{\Eb}{\No}}
\safemath{\EsNo}{\frac{\Es}{\No}}

\DeclareMathOperator{\CHop}{\ensuremath{\opH}} 
\safemath{\tvir}{\rndh_{\CHop}}				
\safemath{\tvtf}{\rndl_{\CHop}}				
\safemath{\spf}{\rnds_{\CHop}}				
\safemath{\bff}{H_{\CHop}}					

\safemath{\ircf}{r_{h}}						
\safemath{\tftvcf}{r_{s}}					
\safemath{\tfcf}{r_{l}}						
\safemath{\bfcf}{r_{H}}						

\safemath{\tcorr}{c_h}						
\safemath{\scf}{c_{s}}						
\safemath{\tfcorr}{c_{l}}					
\safemath{\fcorr}{c_{H}}						

\safemath{\mi}{I}							
\safemath{\capacity}{C}						

\safemath{\normal}{\mathcal{N}}			
\safemath{\jpg}{\mathcal{CN}}			
\safemath{\mchain}{\leftrightarrow}		

\safemath{\dB}{\,\mathrm{dB}}
\safemath{\dBm}{\,\mathrm{dBm}}
\safemath{\Hz}{\,\mathrm{Hz}}
\safemath{\kHz}{\,\mathrm{kHz}}
\safemath{\MHz}{\,\mathrm{MHz}}
\safemath{\GHz}{\,\mathrm{GHz}}
\safemath{\s}{\,\mathrm{s}}
\safemath{\ms}{\,\mathrm{ms}}
\safemath{\mus}{\,\mathrm{\text{\textmu}s}}
\safemath{\ns}{\,\mathrm{ns}}
\safemath{\ps}{\,\mathrm{ps}}
\safemath{\meter}{\,\mathrm{m}}
\safemath{\mm}{\,\mathrm{mm}}
\safemath{\cm}{\,\mathrm{cm}}
\safemath{\m}{\,\mathrm{m}}
\safemath{\W}{\,\mathrm{W}}
\safemath{\mW}{\, \mathrm{mW}}
\safemath{\J}{\,\mathrm{J}}
\safemath{\K}{\,\mathrm{K}}
\safemath{\bit}{\,\mathrm{bit}}
\safemath{\nat}{\,\mathrm{nat}}

\safemath{\define}{\triangleq}			

\safemath{\equivalent}{\sim}
\safemath{\distas}{\sim}					
\safemath{\sdiff}{\Delta}				

\safemath{\reals}{\mathbb{R}}
\safemath{\positivereals}{\reals_{+}}
\safemath{\integers}{\mathbb{Z}}
\safemath{\posint}{\integers_{+}}
\safemath{\naturals}{\mathbb{N}}
\safemath{\posnaturals}{\naturals_{+}}
\safemath{\complexset}{\mathbb{C}}
\safemath{\rationals}{\mathbb{Q}}

\newcommand*{\fancyrefapplabelprefix}{app}		
\newcommand*{\fancyrefthmlabelprefix}{thm}		
\newcommand*{\fancyreflemlabelprefix}{lem}		
\newcommand*{\fancyrefcorlabelprefix}{cor}		
\newcommand*{\fancyrefdeflabelprefix}{def}		
\newcommand*{\fancyrefproplabelprefix}{prop}	
\newcommand*{\fancyrefobslabelprefix}{obs}		
\newcommand*{\fancyrefalglabelprefix}{alg}		
\newcommand*{\fancyrefasmlabelprefix}{asm}	    

\frefformat{vario}{\fancyrefseclabelprefix}{Section~#1}
\frefformat{vario}{\fancyrefthmlabelprefix}{Theorem~#1}
\frefformat{vario}{\fancyreflemlabelprefix}{Lemma~#1}
\frefformat{vario}{\fancyrefcorlabelprefix}{Corollary~#1}
\frefformat{vario}{\fancyrefdeflabelprefix}{Definition~#1}
\frefformat{vario}{\fancyrefobslabelprefix}{Observation~#1}
\frefformat{vario}{\fancyrefasmlabelprefix}{Assumption~#1}
\frefformat{vario}{\fancyreffiglabelprefix}{Fig.~#1}
\frefformat{vario}{\fancyrefapplabelprefix}{Appendix~#1} 
\frefformat{vario}{\fancyrefproplabelprefix}{Proposition~#1}
\frefformat{vario}{\fancyrefalglabelprefix}{Algorithm~#1}
\frefformat{vario}{\fancyrefeqlabelprefix}{(#1)}

\safemath{\dictab}{[\,\dicta\,\,\dictb\,]}

\safemath{\ysig}{\bmy}
\safemath{\ysighat}{\hat{\ysig}}
\safemath{\ysigdim}{M}
\safemath{\xsig}{\bmx}
\safemath{\xsigdim}{N}
\safemath{\nx}{n_x}
\safemath{\zsig}{\bmz}
\safemath{\zsigdim}{\ysigdim}
\safemath{\rsig}{\bmr}
\safemath{\Adict}{\bA}
\safemath{\Adicttilde}{\widetilde{\Adict}}
\safemath{\Adictdim}{\outputdim\times\xsigdim}
\safemath{\avec}{\bma}
\safemath{\avectilde}{\tilde{\avec}}
\safemath{\Bdict}{\bB}
\safemath{\Bdicttilde}{\widetilde{\Bdict}}
\safemath{\Cdict}{\bC}
\safemath{\cvec}{\bmc}
\safemath{\Ddict}{\bD}
\safemath{\Ddictdim}{\ysigdim\times\xsigdim}
\safemath{\dvec}{\bmd}
\safemath{\Ddicttilde}{\widetilde{\bD}}
\safemath{\Bonb}{\bB}
\safemath{\bvec}{\bmb}
\safemath{\Bonbdim}{\ysigdim\times\ysigdim}
\safemath{\noise}{\bmn}
\safemath{\noisedim}{\ysigim}
\safemath{\err}{\bme}
\safemath{\errdim}{\ysigdim}
\safemath{\errset}{\setE}
\safemath{\nerr}{n_e}
\safemath{\delop}{\bP_\errset}
\safemath{\delopc}{\bP_{{\errset}^c}}

\safemath{\cplxi}{\imath}
\safemath{\cplxj}{\jmath}

\safemath{\dict}{\matD}
\safemath{\inputdim}{N}		
\safemath{\outputdim}{M}		
\safemath{\sparsity}{S}	
\safemath{\inputdimA}{{N_a}}	
\safemath{\inputdimB}{{N_b}}	
\safemath{\elemA}{{n_a}}	
\safemath{\elemB}{{n_b}}	
\safemath{\resA}{\matR_a}	
\safemath{\resB}{\matR_b}	
\safemath{\subD}{\matS} 
\safemath{\subA}{\matS_a} 
\safemath{\subB}{\matS_b} 
\safemath{\dicta}{\matA} 	
\safemath{\dictb}{\matB} 	
\safemath{\hollowS}{H}
\safemath{\hollowA}{H_a}
\safemath{\hollowB}{H_b}
\safemath{\cross}{Z}
\safemath{\coh}{\mu_d}			
\safemath{\coha}{\mu_a}			
\safemath{\cohb}{\mu_b}			
\safemath{\mubs}{\nu}	
\safemath{\cohm}{\mu_m} 
\safemath{\dictset}{\setD}	
\safemath{\dictsetp}{\dictset(\coh,\coha,\cohb)}	
\safemath{\dictsetgen}{\dictset_\text{gen}}
\safemath{\dictsetgenp}{\dictsetgen(\coh)}
\safemath{\dictsetonb}{\dictset_\text{onb}}
\safemath{\dictsetonbp}{\dictsetonb(\coh)}

\safemath{\leftside}{U}
\safemath{\rightsideA}{R_a}
\safemath{\rightsideB}{R_b}

\safemath{\indexS}{\setI_S} 

\safemath{\na}{n_a}			
\safemath{\nb}{n_b}			
\safemath{\coeffa}{p_i}	
\safemath{\coeffb}{q_j}	
\safemath{\seta}{\setP}		
\safemath{\setb}{\setQ}     
\safemath{\setw}{\setW}	
\safemath{\setz}{\setZ}	
\safemath{\cola}{\veca}		
\safemath{\colb}{\vecb}		
\safemath{\cold}{\vecd}		
\safemath{\inputvec}{\vecx} 	
\safemath{\error}{\vece}	
\safemath{\noiseout}{\vecz} 	
\safemath{\inputvecel}{x}
\safemath{\inputveca}{\vecx_a}
\safemath{\inputvecb}{\vecx_b}
\safemath{\outputvec}{\vecy}	
\safemath{\lambdamin}{\lambda_{\mathrm{min}}}

\safemath{\elltwo}{\ell_2}
\safemath{\ellone}{\ell_1}
\safemath{\ellzero}{\ell_0}
\safemath{\ellinf}{\ell_\infty}
\safemath{\ellinftilde}{\ell_{\widetilde\infty}}
\safemath{\licard}{Z(\coh,\coha,\cohb)}
\safemath{\xsol}{\hat{x}}
\safemath{\xbord}{x_b}		
\safemath{\xstat}{x_s}		
\safemath{\xstatLone}{\tilde{x}_s}
\safemath{\order}{\mathcal{O}} 
\safemath{\scales}{\Theta} 
\safemath{\ones}{\mathbf{1}} 
\safemath{\zeroes}{\mathbf{0}} 
\safemath{\thlone}{\kappa(\coh,\cohb)} 
\safemath{\constoneA}{\delta} 
\safemath{\constoneB}{\epsilon} 
\safemath{\nlarge}{L}				   
\safemath{\sumlarge}{S_\nlarge}
\safemath{\maxlarger}{P_\nlarge}	   
\safemath{\Pzero}{\textrm{P0}}	
\safemath{\Pone}{\textrm{P1}}
\safemath{\vecfir}{\vecw}			 
\safemath{\vecsec}{\vecz}
\safemath{\elvecfir}{w}              
\safemath{\elvecsec}{z}				 
\safemath{\nlargefir}{n}
\safemath{\normout}{\gamma}
\safemath{\auxfun}{h}
\safemath{\supp}{\textrm{supp}}

\safemath{\indexa}{\ell}
\safemath{\indexb}{r}
\safemath{\indexc}{i}
\safemath{\indexd}{j}

\safemath{\project}{P}

\title{Soft-Output Finite Alphabet Equalization for mmWAVE Massive  MIMO}
\name{Oscar Casta\~neda$^1$, Sven Jacobsson$^{2,3}$, Giuseppe Durisi$^3$, Tom Goldstein$^4$, and Christoph Studer$^1$\thanks{The work of OC was supported in part by ComSenTer, a Semiconductor Research Corporation (SRC) program, by SRC nCORE task 2758.004, and by a Qualcomm Innovation Fellowship. The work of SJ and GD was supported by the Swedish Foundation for Strategic Research under grant ID14-0022, and by the Swedish Governmental Agency for Innovation Systems (VINNOVA). The work of TG was supported in part by the US NSF under grant CCF-1535902 and by the US Office of Naval Research under grant N00014-17-1-2078. The work of CS was supported in part by Xilinx Inc.\ and by the US NSF under grants ECCS-1408006 and CCF-1535897.}\thanks{The present work extends its journal version \cite{castaneda19fame} by providing an unbiased equalizer with soft-output capabilities as well as results for a coded mmWave massive MU-MIMO system.}}
\address{$^1$Cornell Tech, New York, NY; e-mail: oc66@cornell.edu, studer@cornell.edu\\
$^2$Ericsson Research, Gothenburg, Sweden; e-mail: sven.jacobsson@ericsson.com\\
$^3$Chalmers University of Technology, Gothenburg, Sweden; e-mail: durisi@chalmers.se\\
$^4$Department of CS, University of Maryland, College Park, MD; e-mail: tomg@cs.umd.edu}
\begin{document}
%
\maketitle
\begin{abstract}
Next-generation wireless systems are expected to combine millimeter-wave (mmWave) and massive multi-user multiple-input multiple-output (MU-MIMO) technologies to deliver high data-rates.
These technologies require the basestations (BSs) to process high-dimensional data at extreme rates, which results in high power dissipation and system costs.
Finite-alphabet equalization has been proposed recently to reduce the power consumption and silicon area of uplink spatial equalization circuitry at the BS by coarsely quantizing the equalization matrix.
In this work, we improve upon finite-alphabet equalization by performing unbiased estimation and soft-output computation for coded systems.
By simulating a massive MU-MIMO system that uses orthogonal frequency-division multiplexing and per-user convolutional coding, we show that soft-output finite-alphabet equalization delivers competitive error-rate performance using only $1$ to $3$ bits per entry of the equalization matrix, even for challenging mmWave channels.
\end{abstract}


%
\vspace{-0.15cm}
\section{Introduction}
\label{sec:intro}
\vspace{-0.15cm}

Future wireless communication systems are likely to combine millimeter-wave (mmWave) communication~\cite{SwindlehurstCommMag} with massive multi-user multiple-input multiple-output (MU-MIMO)~\cite{larsson14-02a} as they enable one to serve multiple user equipments (UEs) simultaneously in the same frequency band with high throughput.
The extreme bandwidths offered at mmWave frequencies combined with the strong path loss,  however, require the deployment of hundreds of antennas at the basestation (BS) and computationally complex baseband processing circuitry.
Consequently, power and system costs are key concerns for designing mmWave MU-MIMO systems in practice.

In order to keep power consumption of MU-MIMO systems within reasonable bounds, energy-efficient hybrid analog-digital solutions~\cite{roh14,sadhu17,alkhateeb14b} have been proposed in the past.
Such hybrid approaches are, however, limited in their ability to capture and resolve multiple arriving signal paths~\cite{alkhateeb14b,bjornson19mimommwave,dutta19}, which degrades spectral efficiency.
In contrast, all-digital BS architectures~\cite{mo15CAOQ,Roth17Digital,jacobssonTAMMU17} are able to overcome this issue, but are commonly perceived as energy inefficient.
Recent results~\cite{dutta19,Roth17Digital} have demonstrated that, by reducing the resolution of the data converters, the power consumption of radio-frequency (RF) circuitry and data converters in all-digital BS architectures is comparable to hybrid solutions.
However, the power consumption and system costs of baseband processing in all-digital BS architectures are largely unexplored. 

\subsection{Finite Alphabet Equalization}
In the uplink (UEs transmit to BS), all-digital spatial equalization is required to recover the signals transmitted by each of the~$U$ UEs from the data converters at the $B$ BS antennas.
Spatial equalization performs complex-valued matrix-vector products between a $U\times B$ equalization matrix and a $B$-dimensional received vector at the rate of the incoming samples. 
For a system with $B=256$ BS antennas and $U=16$ UEs, performing a single matrix-vector product at a rate of $2$\,G\,samples/s consumes already $28$\,W and  $129\,\text{mm}^2$ of silicon area in $28$\,nm CMOS~\cite{castaneda19fame}.
For wideband systems that use orthogonal frequency-division multiplexing (OFDM), these power and area numbers are expected to increase even further.
Clearly, efficient spatial equalization circuitry is necessary to lower the power consumption and silicon area of all-digital BS architectures, without hampering their spectral efficiency.

The power consumption and silicon area of matrix-vector products can be decreased by reducing the bit resolution of their constituent multiplications and additions. 
Existing work has mainly focused on the use of low-resolution (e.g., $1$ to $8$ bits) data converters at the BS antennas of massive MU-MIMO systems~\cite{alkhateeb14b,dutta19,mo15CAOQ,Roth17Digital,studer16a,yan19},
which reduces the precision of the received vectors.
However, even when using low-resolution vectors, the equalization matrix is typically represented with high-resolution numbers, e.g., $10$ to $12$ bits~\cite{studer2011asic,WYWDCS2014}.
In the recent work~\cite{castaneda19fame}, we proposed \emph{finite-alphabet equalization}, a novel paradigm that uses low-resolution numbers to represent the entries of the equalization matrices.
To mitigate the loss in performance caused by low-precision equalization matrices, we introduced the finite-alphabet minimum mean-square error equalization (FAME) problem in~\cite{castaneda19fame}. This new approach enables the computation of low-precision equalization matrices that minimize the post-equalization mean-square error (MSE).

\subsection{Contributions}
In this paper, we extend finite-alphabet equalization as put forward in~\cite{castaneda19fame} by unbiased estimation and soft-output computation.
We derive a compact expression of the  post-equalization MSE, which can be used to efficiently compute log-likelihood ratio (LLR) values. 
We demonstrate the effectiveness of our methods by providing error-rate simulation results for a coded massive MU-MIMO-OFDM system, for two unbiased soft-output finite-alphabet equalizers, both in line-of-sight (LoS) and non-LoS mmWave channel scenarios.

\subsection{Notation}
Uppercase and lowercase boldface letters denote matrices and column vectors, respectively. 
For a matrix $\bA$, the Hermitian transpose is $\bA^H$, the Frobenius norm is $\|\bA\|_{F}$, the real part is $\Re\{\bA\}$, and the imaginary part is $\Im\{\bA\}$. 
$\bI_M$ is the $M\times M$ identity matrix.
For a vector~$\bma$, the $k$th entry is $a_k$, the  $\ell_2$-norm is $\vecnorm{\veca}_2$, and the entry-wise complex conjugate is $\bma^*$. 
The $k$th standard basis vector is $\bme_k$.
The signum function~$\text{sgn}(a)$ returns~$+1$ for~$a\ge0$ and~$-1$ otherwise.
$\Ex{\bmx}{\cdot}$ is the expectation operator with respect to the random vector $\bmx$.

\section{System Model and Equalization}
\label{sec:prereq}

\subsection{Uplink System Model}
\begin{figure}[tp]
\centering
\includegraphics[width=0.9\columnwidth]{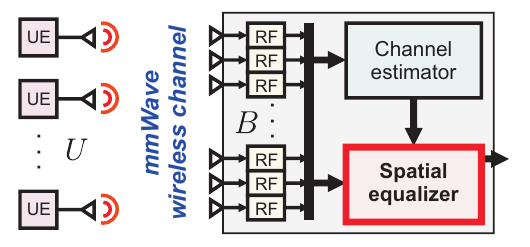}
\caption{Uplink of a massive MU-MIMO mmWave system. The $U$ UEs transmit data to the $B$-antenna BS. After estimating the channel, the all-digital BS uses spatial equalization to recover the UEs' individual signals. Finite-alphabet equalization \cite{castaneda19fame} consists of using low-resolution spatial equalization matrices.}
\label{fig:system_overview}
\end{figure}
As illustrated in \fref{fig:system_overview}, we focus on a massive MU-MIMO system where $U$ single-antenna UEs transmit data to a BS with  $B$ antennas.
The system uses OFDM with $W$ subcarriers, where the frequency-domain input-output relation per subcarrier $w\in\{1,\ldots,W\}$ is $\bmy_w=\bH_w\bms_w+\bmn_w$.
At subcarrier $w$, $\bmy_w\in\complexset^B$ is the vector received at the BS,  $\bH_w\in\complexset^{B\times U}$ is the uplink MIMO channel matrix, $\bms_w\in\setS^U$ is the transmit data vector, where $\setS$ is the constellation (e.g., 16-QAM),  and  $\bmn_w\in\complexset^B$ is i.i.d.\ circularly-symmetric complex Gaussian noise with covariance matrix $\bC_{\bmn_w}=\Ex{\bmn_w}{\bmn_w\bmn_w^H}=\No\bI_B$.
To simplify notation, we will omit the subcarrier index $w$ and focus (without loss of generality) on a single subcarrier.
We assume that the transmit signals $s_u$, $u=1,\ldots,U$, of the UEs are i.i.d.\ with zero mean and variance $\Es$; this ensures that $\bC_\bms = \Ex{\bms}{\bms\bms^H}=\Es\bI_U$.
We also assume that the channel remains constant over several symbol transmissions, so that the BS is able to estimate the channel matrix---for simplicity, we assume perfect channel state information at the~BS. 

\subsection{Unbiased L-MMSE Equalization}
A central task at the BS is to generate estimates of the transmit data vector $\bms$ using the received vector $\bmy$ and knowledge of the channel matrix $\bH$. 
At the high bandwidths offered by mmWave systems, linear estimators are preferable due to their simplicity. 
We therefore focus on linear spatial equalizers that compute estimates $\bar\bms$ of the transmit signals $\bms$ as $\bar\bms=\bW^H \bmy$.
Here, $\bW^H\in\complexset^{U\times B}$ is the linear minimum MSE~\mbox{(L-MMSE)} equalization matrix, which minimizes the MSE defined by 
\begin{align} \label{eq:MSE}
\textit{MSE} = \Ex{\bms,\bmn}{\|\bar\bms-\bms\|^2_2}\!.
\end{align}
Under the statistical assumptions on $\bms$ and $\bmn$ listed above, the L-MMSE equalization matrix is given by \cite{paulraj03}
\begin{align} \label{eq:LMMSEmatrix}
\bW^H = (\rho\bI_U+\bH^H\bH)^{-1}\bH^H,
\end{align}
where $\rho=\No/\Es$.
The rows~$\bmw_u^H $, $u=1,\ldots,U$, of the L-MMSE equalizer $\bW^H$ can be computed by solving
\begin{align} \label{eq:MMSEequalizer}
\bmw_u  = \argmin_{\tilde\bmw\in\complexset^{B}} \|\bme_u-\bH^H\tilde\bmw\|_2^2 + \rho\|\tilde\bmw\|^2_2.
\end{align}

Spatial equalization with the biased L-MMSE estimate for each user $u=1,\ldots,U$ amounts to computing  
\begin{align}
\bar{s}_u= \bmw_u^H\bmy = \bmw_u^H\bmh_u s_u + \bmw_u^H\tilde\bmn_u,
\end{align}
where~$\bmh_u$ is the $u$th column of $\bH$ and $\tilde\bmn_u =\sum_{i=1,i\neq u}^{U} \bmh_i s_i + \bmn$ is the noise-plus-interference (NPI) vector. 
In general, the L-MMSE equalizer has rows for which $\bmw_u^H\bmh_u \neq 1$.
Thus, to perform \emph{unbiased} estimation, our goal is to compute the estimates for each UE $u=1,\ldots,U$ as follows:
\begin{align} \label{eq:unbiasedeq}
\hat{s}_u=\frac{\bar{s}_u}{\bmw_u^H\bmh_u}=\frac{\bmw_u^H\bmy}{\bmw_u^H\bmh_u}=s_u+\frac{\bmw_u^H\tilde\bmn_u}{\bmw_u^H\bmh_u}.
\end{align}
In general, the  biased $\bar{s}_u$ and unbiased $\hat{s}_u$ estimates differ: Biased estimates minimize the MSE in \fref{eq:MSE}, whereas unbiased estimates typically achieve lower error rates \cite{studer2011asic}. 

%

%
\section{Finite-Alphabet Equalization}
\label{sec:fa_eq}
For the high dimensions and data rates generated by mmWave massive MU-MIMO-OFDM systems, spatial equalization with matrix-vector products (e.g., using $\bar\bms=\bW^H\bmy$) leads to power-hungry circuitry and large silicon area.
To lower power and area, finite-alphabet equalization, proposed in~\cite{castaneda19fame}, uses low-resolution numbers to represent the entries of $\bW^H$, which enables the use of low-power and low-area multipliers and adders.
Unfortunately, a na\"ive quantization of the entries of the L-MMSE matrix $\bW^H$ would result in a significant error-rate performance degradation.
To mitigate this issue while still being able to reduce hardware complexity, we proposed in~\cite{castaneda19fame} to use finite-alphabet equalization matrices described next.

\subsection{Unbiased Finite-Alphabet  Equalization}
As defined in \cite{castaneda19fame}, a \emph{finite-alphabet equalization matrix} is a $U\times B$ matrix that has the following form:
\begin{align} \label{eq:FAMEmatrix}
\bV^H = \mathrm{diag}(\boldsymbol\beta^*) \bX^H.
\end{align}
Here, the vector $\boldsymbol\beta\in\complexset^U$ contains post-equalization scaling factors and $\bX^H\in\setX^{U\times B}$ is a \emph{low-resolution} equalization matrix with entries taken from a \emph{low-cardinality} finite alphabet~$\setX$.
By applying the structure of \fref{eq:FAMEmatrix} to the equalization procedure in \fref{eq:unbiasedeq}, per-user \emph{unbiased} equalization corresponds to
\begin{align} \label{eq:equalization}
\hat{s}_u=\frac{\bmv_u^H\bmy}{\bmv_u^H\bmh_u}=\frac{\beta_u^*\bmx_u^H\bmy}{\beta_u^*\bmx_u^H\bmh_u}=\frac{\bmx_u^H\bmy}{\bmx_u^H\bmh_u}.
\end{align}
Here, $\bmv^H_u\in\complexset^{1\times B}$ and $\bmx^H_u\in \setX^{1\times B}$ are the $u$th rows of $\bV^H$ and $\bX^H$, respectively. 
Although unbiased equalization as in~\fref{eq:equalization} differs from biased equalization $\beta_u^*\bmx_u^H\bmy$ as originally proposed in~\cite{castaneda19fame}, we emphasize that \fref{eq:equalization}, as its biased counterpart, also reduces hardware complexity.
Concretely, the inner product $\bmx_u^H\bmy$ (formed by $B$ scalar products) can be computed with low-resolution multipliers and adders.
The resulting inner product is then scaled by $1/(\bmx_u^H\bmh_u)$, which requires only one high-resolution scalar multiplication per user.

\subsection{FAME: Finite-Alphabet MMSE Equalization}
To compute finite-alphabet equalization matrices that minimize the post-equalization MSE as in \fref{eq:MSE}, the work in~\cite{castaneda19fame} formulates the FAME problem.
Analogous to \fref{eq:MMSEequalizer}, the rows $\bmv_u^H=\beta_u^*\bmx_u^H$, $u=1,\ldots,U$, of a FAME matrix are obtained by solving
\begin{align} \label{eq:FAME}
\{\beta_u,\bmx_u\} = \argmin_{\tilde\bmx\in\setX^B, \, \tilde\beta\in\complexset}\|\bme_u- \bH^H\tilde\beta\tilde\bmx\|_2^2 + \rho \|\tilde\beta\tilde\bmx\|_2^2.
\end{align}
For a fixed $\beta_u$, the FAME problem in \fref{eq:FAME} is NP-hard \cite{agrell02a,fincke85a}.
To develop practical algorithms, reference~\cite{castaneda19fame} reformulates the problem in \fref{eq:FAME} using a two-step procedure: First, compute
\begin{align} \label{eq:FAMEcompact}
\bmx_u = \argmin_{\tilde\bmx\in\setX^B} \frac{\|\bH^H\tilde\bmx\|_2^2+\rho\|\tilde\bmx\|_2^2}{|\bmh^H_u\tilde\bmx|^2}.
\end{align}
Then, extract the scaling factor $\beta_u(\bmx_u)$ using
\begin{align} \label{eq:optimalscaling}
\beta_u(\vecx_u)
= \frac{ \bmx^H_u\bmh_u }{\|\bH^H\bmx_u\|_2^2 + \rho\|\bmx_u\|_2^2}.
\end{align}
This formulation can be used to derive approximate, low-complexity algorithms; see \fref{sec:fa_alg} for more details.  

\subsection{Soft-Output Finite-Alphabet Equalization}
While the paper~\cite{castaneda19fame} focuses on hard-output data detection, coded communication systems benefit from spatial equalizers that compute soft-outputs. 
To fully exploit forward error correction, we  first extract the post-equalization NPI variance, which is then used to generate LLR values.
For the $u$th UE, the NPI variance is given by the MSE of the unbiased estimate~$\hat{s}_u$, which is computed as follows:
\begin{align}
\nu^2_u &  = \Ex{\bms,\bmn}{|\hat{s}_u-s_u|^2} \\
& \overset{(a)}{=} \frac{\Ex{\bms,\bmn}{\left\lvert\bmx_u^H\bH(\bI_U-\bme_u\bme_u^H)\bms+\bmx_u^H\bmn\right\rvert^2}}{\left\lvert\bmx_u^H\bmh_u\right\rvert^2} \\
& = \frac{\Es\left(\|\bH^H\bmx_u\|_2^2-|\bmx_u^H\bmh_u|^2\right)
+\No\|\bmx_u\|_2^2}{\left\lvert\bmx_u^H\bmh_u\right\rvert^2} \\
& = \frac{\Es}{\bmh_u^H\bmx_u}\frac{\|\bH^H\bmx_u\|_2^2
+\rho\|\bmx_u\|_2^2}{\bmx_u^H\bmh_u}-\Es \\
  & \overset{(b)}{=} \Es\left((\beta_u(\bmx_u)\bmh_u^H\bmx_u)^{-1}-1\right)\!. \label{eq:lastline}
\end{align}
Here, $(a)$ follows from \fref{eq:unbiasedeq} and $(b)$ from \fref{eq:optimalscaling}.
Note that this result applies to any finite-alphabet equalizer as in \fref{eq:FAMEmatrix}, as long as $\beta_u(\bmx_u)$ is computed as in \fref{eq:optimalscaling}.
With this, we can compute soft outputs in the form of LLR values, by assuming that the residual error $\hat{s}_u-s_u$ is circularly-symmetric Gaussian with variance $\nu^2_u$.  Concretely, we compute the  LLR values as follows \cite{tuchler02,studer2011asic}:
\begin{align} 
\Lambda_{u,q} =\, & \textstyle  \log\!\left( \sum_{s\in\setS^{(1)}_q}  \exp\!\left(-\frac{|\hat{s}_u-s|^2}{\nu^2_u}\right) \right) \nonumber \\
& \textstyle  - \log\!\left( \sum_{s\in\setS^{(0)}_q}  \exp\!\left(-\frac{|\hat{s}_u-s|^2}{\nu^2_u}\right) \right)\!. \label{eq:llrvalue}
\end{align}
Here, $\setS^{(1)}_q$ and $\setS^{(0)}_q$ are the subsets of the constellation $\setS$ in which the $q$th bit is $1$ and $0$, respectively.
We note that computing soft outputs for finite-alphabet equalizers entails the same complexity as for infinite-precision L-MMSE \cite{studer2011asic}.
%


\begin{figure*}[tp]
\centering
\subfigure[i.i.d. Rayleigh]{\includegraphics[width=.64\columnwidth]{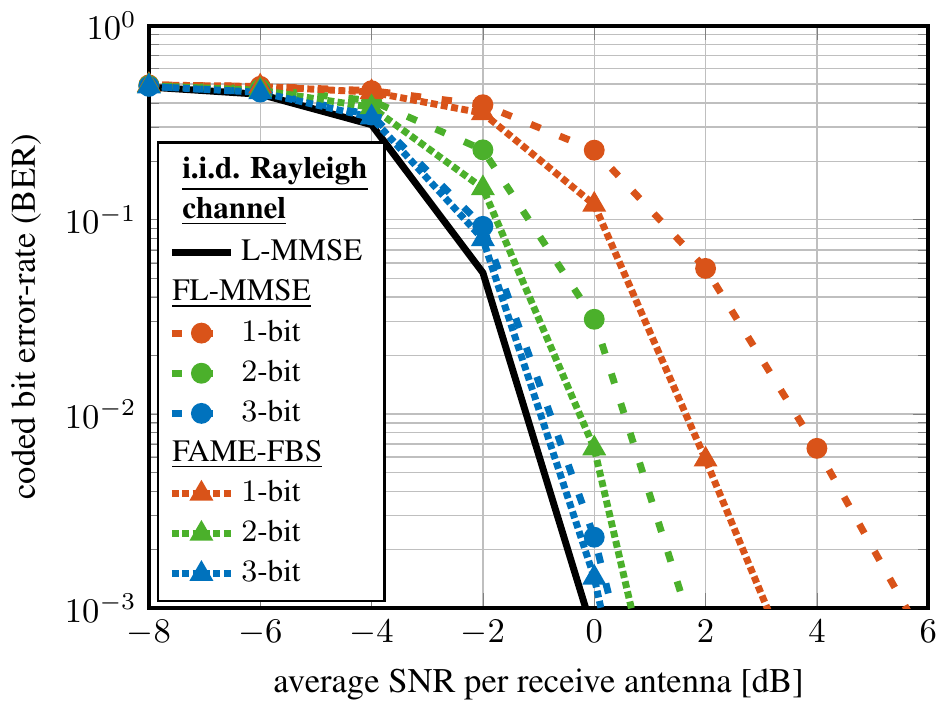}\label{fig:ber_rayleigh}}
\hfill
\subfigure[QuaDRiGa non-LoS]{\includegraphics[width=.64\columnwidth]{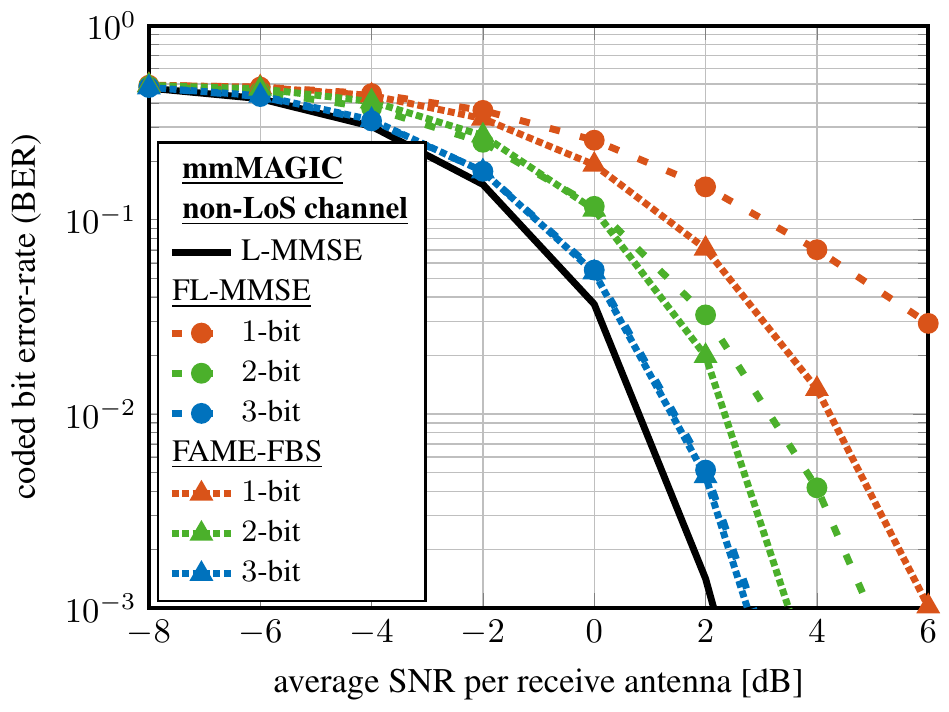}\label{fig:ber_nlos}}
\hfill
\subfigure[QuaDRiGa LoS]{\includegraphics[width=.64\columnwidth]{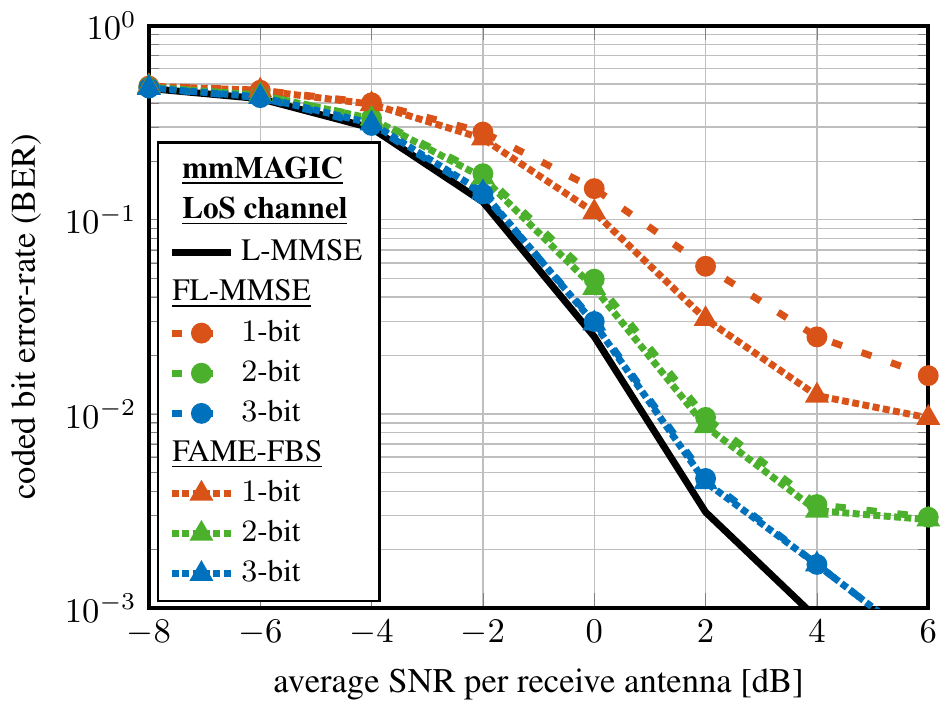}\label{fig:ber_los}}
\vspace{-0.2cm}
\caption{BER of a $B=256$ BS antenna, $U=16$ UE, $16$-QAM, rate-$3/4$ coded OFDM system. FAME-FBS runs $t_\text{max}\le5$ iterations for all but the $1$-bit non-Rayleigh and $2$-bit non-LoS cases, for which $t_\text{max}=20$. For these last scenarios and the Rayleigh case, FAME-FBS is initialized with the MRC equalizer $\bH^H$; otherwise, it is initialized with the FL-MMSE $\bX^H$.}\label{fig:ber_all}
\vspace{-0.33cm}
\end{figure*}

\section{Computing Finite-Alphabet Equalizers}
\label{sec:fa_alg}
We now summarize two algorithms put forward in~\cite{castaneda19fame} to obtain the rows $\bmx_u^H$ of $\bX^H$ in \fref{eq:FAMEmatrix}.
For both algorithms, once $\bmx_u^H$ is known, the associated $\beta_u(\bmx_u)$ is computed using \fref{eq:optimalscaling}; this factor is required to compute the variance $\nu_u^2$ using \fref{eq:lastline}, which is then used to compute LLR values with \fref{eq:llrvalue}.

\subsection{Finite-Alphabet L-MMSE (FL-MMSE)}
As described in~\cite{castaneda19fame}, a simple way of computing finite-alphabet equalization vectors $\bmx_u^H$ is to take a row of the infinite-precision L-MMSE equalizer $\bmw_u^H$ in \fref{eq:LMMSEmatrix} and quantize its entries to low resolution. 
Note that this approach, dubbed FL-MMSE, does not need to solve the FAME problem in~\fref{eq:FAMEcompact}.
To quantize a row of the L-MMSE matrix $\bmw_u^H$, we follow the procedure described in \cite{castaneda19fame}:
We find the maximum-magnitude entry $w_\text{max}$ in $[\Re\{\bmw_u^H\},\Im\{\bmw_u^H\}]$, and quantize the entries of $\Re\{\bmw_u^H\}$ and $\Im\{\bmw_u^H\}$ using uniform-width bins across the range $[-w_\text{max},w_\text{max}]$.
Each bin is represented by its centroid value.
Then, the centroid values are scaled by the same factor so that all of them are integers, which can be  represented in hardware using few bits.
Note that such scaling does not affect the value of the objective function in \fref{eq:FAMEcompact}.

\subsection{FAME via Forward-Backward Splitting (FBS)}
As detailed in \cite{castaneda19fame}, the FAME problem in \fref{eq:FAMEcompact} can be approximately solved with FBS \cite{BT09,GSB14}, resulting in the FAME-FBS procedure executed for $t=1,\ldots,t_\text{max}$ iterations:
\begin{align}
\tilde\bmz^{(t+1)} & = \left(\bI_B - \tau^{(t)}\bH(\bI_U-\gamma^{(t)}\bme_u\bme_u^H)\bH^H\right) \tilde\bmx^{(t)} \label{eq:step1}\\
\tilde\bmx^{(t+1)} & = \mathrm{proj} (\tilde\bmz^{(t+1)}). \label{eq:step2}
\end{align}
The proximal operator $\text{proj}(\tilde{z})=\text{sgn}\left(\lbrace\tilde{z}\rbrace\right)\min\left\{\nu^{(t)}|\lbrace\tilde{z}\rbrace|,1\right\}$ is applied element-wise and separately to $\Re\{\tilde\bmz\}$ and $\Im\{\tilde\bmz\}$.
Here, {$\{\tau^{(t)}\}$}, {$\{\nu^{(t)}\}$}, and {$\{\gamma^{(t)}\}$} are per-iteration parameter sets that are   tuned empirically.
This iterative process can be initialized with the maximum ratio-combining (MRC) equalizer $\tilde\bmx^{(1)}=\bmh_u$ or with the result $\bmx_u$ of  FL-MMSE.
The output of the final iteration, $\tilde\bmx^{(t_\text{max}+1)}$, is quantized to the finite-alphabet set $\setX$ by using the same approach as for FL-MMSE but with $w_\text{max}=1$ for all UEs.

\section{Simulation Results}
\fref{fig:ber_all} shows coded bit error-rate (BER) for FL-MMSE and FAME-FBS using $1$ to $3$ bits per real and imaginary part for each entry of the low-resolution equalization matrix $\bX^H$.
The simulation results correspond to a $B=256$ BS antenna, $U=16$ UE, $16$-QAM system, with OFDM transmission over $W=1200$ subcarriers. 
 We use per-UE rate-$3/4$ convolutional codes and soft-input Viterbi decoding. 
The BER curves are obtained 
for three propagation conditions: (a) Rayleigh fading, (b) non-LoS, and (c) LoS.
To model mmWave systems, the non-LoS and LoS channels are obtained using the QuaDRiGa model~\cite{jaeckel2014quadriga} with the ``mmMAGIC\_UMi'' scenario; we consider a uniform linear array with half-wavelength antenna spacing and transmission at a carrier frequency of $60$\,GHz. 
Each subcarrier has a bandwidth of $240$\,kHz and power control ensures a $\pm3$\,dB power variation among UEs.

From \fref{fig:ber_all}, we see that the coded  BER performance of FAME-FBS meets or exceeds that of FL-MMSE for all of the considered scenarios.
The discrepancy between these two methods decreases when increasing the number of bits used for the finite-alphabet equalization matrix.
While with $1$-bit, FAME-FBS offers more than $10\times$ lower BER at $6$\,dB SNR compared to FL-MMSE for the non-LoS channel, the performance of the $3$-bit FL-MMSE and FAME-FBS is practically the same and approaches that of the infinite-precision L-MMSE by less than $1.5$\,dB for all considered scenarios.
%

\section{Conclusions}
\label{sec:conc}

We have extended the finite-alphabet equalization paradigm introduced in~\cite{castaneda19fame}.
Specifically, we have proposed an unbiased soft-output finite-alphabet equalizer that can be used in coded communication systems.
We have derived a post-equalization MSE expression that can be computed efficiently and is used to  compute LLR values. 
Simulation results for a coded mmWave massive MU-MIMO-OFDM system have shown that finite-alphabet equalization delivers a competitive error-rate that approaches that of the infinite-precision L-MMSE equalizer, so much as virtually reaching it with as few as $3$ bits, even for realistic mmWave channels.
These results pave the way for all-digital BS architectures that reduce power consumption and silicon area, while preserving high spectral efficiency.
\bibliographystyle{IEEEtran}
\bibliography{IEEEabrv,confs-jrnls,publishers,svenbib,studer}

\end{document}